\documentclass{article}
\usepackage{dcase2018,amsmath,graphicx,url,times,booktabs, tabularx}
\usepackage{algorithm, algpseudocode}

\usepackage{xcolor}
\usepackage{multirow,multicol}

\definecolor{monvert}{rgb}{0,0.8,0.5}
\title{Large-scale weakly labeled semi-supervised sound event detection in domestic environments}

\name{Romain Serizel$^{1}$,
      Nicolas Turpault$^{1}\sthanks{This work has been funded by the French region Grand-Est.}$,
      Hamid Eghbal-Zadeh$^{2}\sthanks{This work has been funded by the Austrian Ministries BMVIT and BMWFW, and the Province of Upper Austria, via the COMET Center SCCH.}$,
      Ankit Parag Shah$^{2}$
      }
\address{$^1$Université de Lorraine, CNRS, Inria, Loria, F-54000 Nancy, France\\
        $^2$Institute of Computational Perception, Johannes Kepler University of Linz, Austria\\
        $^3$Language Technologies Institute, Carnegie Mellon University, Pittsburgh PA, United States\\
 }

\begin{document}

\ninept
\maketitle

\begin{sloppy}

\begin{abstract}
This paper presents DCASE 2018 task 4. The task evaluates systems for the large-scale detection of sound events using weakly labeled data (without time boundaries). The target of the systems is to provide not only the event class but also the event time boundaries given that multiple events can be present in an audio recording. Another challenge of the task is to explore the possibility to exploit a large amount of unbalanced and unlabeled training data together with a small weakly labeled training set to improve system performance. The data are Youtube video excerpts from domestic context which have many applications such as ambient assisted living. The domain was chosen due to the scientific challenges (wide variety of sounds, time-localized events\dots) and potential industrial applications.
\end{abstract}


\section{Introduction}
\label{sec:intro}
We are constantly surrounded by sounds and we rely heavily on these sounds to obtain important information about what is happening around us. Ambient sound analysis aims at automatically extracting information from these sounds. It encompasses disciplines such as sound scene classification (in which context does this happen?), sound event detection (SED) and classification (what happens during this recording?)~\cite{virtanen2018computational}. It has been attracting a continuously growing attention during the past years as it can have a great impact in many applications including smart cities, autonomous cars or ambient assisted living.

In this task, we focus on SED with time boundaries in domestic applications. The system then has to detect when an sound event occurs and what is the class of the event (as opposed to audio tagging where only the presence of a sound event is important regardless of when it happened). Current systems heavily rely on a supervised training phase and usually require a large set of sound recordings labeled in terms of event with time boundaries. Obtaining such annotations is tedious and it is hardly feasible to gather a sufficient amount of data to train state-of-the-art systems that are often relying on complex deep networks architectures~\cite{parascandolo_recurrent_2016, takahashi_deep_2016, adavanne_sound_2017, cakir_convolutional_2017, xu_large-scale_2017}.

We propose to follow-up on DCASE2017 task 4~\cite{mesaros2017dcase} and investigate the scenario where a large scale corpus is available but only a small amount of the data is labeled. We propose to use a subset of the Audioset corpus~\cite{audioset} targeting classes of sound events related to domestic applications. The labels are provided at clip level (an event is present or not within a sound clip) but without the time boundaries (weak labels, that can also be refereed to as tags) in order to further decrease the annotation time. These constraints indeed corresponds to constraints faced in many real applications where the budget allocated to annotation is limited.

In order to fully exploit this dataset, the proposed systems will have to tackle two different problems. The first problem is related to the exploitation of the unlabeled part of the dataset either in unsupervised approaches~\cite{Salamon, Jansen} or together with the labeled subset in semi-supervised approaches~\cite{Zhang2012,komatsu2016acoustic,ankitsemisupervised}. The second problem is related to the detection of the time boundaries and how to train a system that can detect these boundaries from weakly labeled data. Currently, most of the state-of-the-art approaches rely mainly on smoothing techniques to ensure time consistency~\cite{xu_large-scale_2017, Jeong2017, Lee2017a} which is not sufficient to estimate the time boundaries accurately. The evaluation metric chosen is penalizing these boundary estimation errors heavily in order to emphasize this latter aspect. 

This manuscript describes DCASE2018 task 4 and is organized as follows. Section~\ref{sec:dataset} presents the dataset used in task 4. The task evaluation procedure is described in Section~\ref{sec:task}. The baseline system is described and evaluated in~\ref{sec:baseline}. Conclusions and perspectives are presented in Section~\ref{sec:conc}

\section{Audio dataset}
\label{sec:dataset}

The task employs a subset of Audioset~\cite{audioset}. Audioset consists of an expanding ontology of 632 classes of sound events and a collection of 2 million human-labeled 10-seconds sound clips (less than 21\% are shorter than 10 seconds) drawn from 2 million YouTube videos. The ontology is specified as a hierarchical graph of event categories, covering a wide range of human and animal sounds, musical instruments and genres, and common everyday environmental sounds.

Audioset provides labels at clip level (without time boundaries for the events) also known as weak labels. Audio clips are collected from Youtube videos uploaded by independent users so the number of clips per class varies dramatically and the dataset is not balanced \footnote{see also \url{https://research.google.com/Audioset//dataset/index.html}}. Google researchers conducted a quality assessment task where experts were exposed to 10 randomly selected clips for each class of sound events and discovered that in most of the cases not all the clips contains the event related to the given label. More information about the initial annotation process can be found in Gemmeke et al.~\cite{audioset}.

We will focus on a subset of Audioset that consists of 10 classes of sound events (Table~\ref{tab:dset}). The development set provided for task 4 is split into a training set and a test set\footnote{The annotations files and the script to download the audio files is available on the git repository for task 4 \url{https://github.com/DCASE-REPO/dcase2018_baseline/tree/master/task4/dataset}}.

\begin{table}
  \centering
\resizebox{0.5\textwidth}{!}{\begin{tabular}{l||c|c|c||c|c}
  \multirow{ 3}{*}{Class}&\multicolumn{3}{c||}{Count}&\multicolumn{2}{c}{Duration (in~s)}\\
&Training&\multicolumn{2}{c||}{Test}&\multicolumn{2}{c}{Test}\\
 & clips& clips& events& mean&median\\
\hline\hline
Alarm/bell/ringing &205 & 45 &	112&1.53&0.58\\
Blender &134&30 &	40&5.35&4.59\\
Cat &173& 32 &97&0.81&0.71\\
Dishes &184&35&	122&0.56&0.42\\
Dog &214&29&127&1.03&0.66\\
Electric shaver/toothbrush &103&25&	28&7.41&8.78\\
Frying &171&24&	24&9.34&10.00\\
Running water &343&63&	76&5.61&5.53\\
Speech &550&105&261&1.51&1.20\\
Vacuum cleaner &167&35&	36&8.66&9.99\\
\hline\hline
Total &2244&288&	906&2.41&1.03\\
\end{tabular}}
\caption{Class-wise statistics}
\label{tab:dset}
\end{table}
\begin{table}
  \centering
\begin{tabular}{l||c|c|c}
Number of classes & 1 & 2 & 3 and more\\\hline
Clip proportion & 62.36\%& 32.89\%& 4.75\%\\
\end{tabular}
\caption{Proportion of clips with multiple classes of sound events (training set)}
\label{tab:train_overlap_seg}
\end{table}

\subsection{Training set}
In order to reflect what could possibly happen in a real-world scenario, we provide 3 different splits of training data in task 4 training set: a labeled training set, an unlabeled in domain training set and an unlabeled out of domain training set.

\subsubsection{Labeled training set}
This set contains 1,578 clips (2,244 class occurrences) for which weak labels provided in Audioset have been verified and corrected by human annotators. The weak labels are provided in a tab-separated csv file under the following format:
\begin{center}
    {\sf [filename (str)][tab][class\_label (str)]}
\end{center}
The first column is the name of the sound clip downloaded from YouTube composed of the YouTube ID of the video and the time boundaries of the 10-seconds audio clip extracted from the video. The last column corresponds to the sound events that are present in the clip, each separated by a semi-colon.

The amount of clips per class of sound events is presented in Table~\ref{tab:dset} and the number of classes observed per clip is presented in Table~\ref{tab:train_overlap_seg}. One-third of the clips in this set contain at least two different classes of sound events.


\subsubsection{Unlabeled in domain training set}
This set contains 14,412 clips. The clips are selected such that the distribution per class of sound event (based on Audioset labels) is close to the distribution in the labeled set. Note however that given the uncertainty on Audioset labels this distribution might not be exactly similar. Audioset labels have not been verified for this subset and should not be used during the systems development.

\subsubsection{Unlabeled out of domain training set}
This set is composed of 39,999 clips extracted from classes of sound events that are not considered in the task. Note that these clips are chosen based on the Audioset labels which are not verified for this subset and therefore might be noisy. Additionally, as the speech class is present in about half of the clips in Audioset, the unlabeled out of domain set also contains almost 20000 clips with speech. This was the only way to have a set which is somehow representative of Audioset. Indeed, discarding speech would have also meant discarding many other classes of sound events and the variability of the set would have been penalized. Audioset labels have not been verified for this subset and should not be used during the systems development.

\subsection{Test set}
\begin{figure*}
  \centering
  \includegraphics[width=\textwidth]{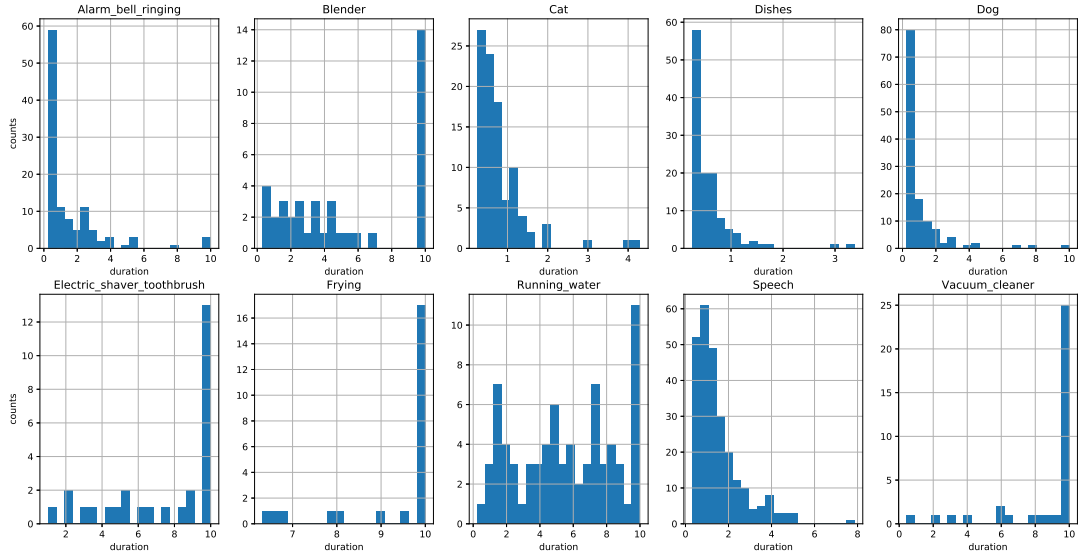}
  \caption{Duration distribution by class of sound events.}
  \label{dur_dist}
\end{figure*}

\begin{table}
  \centering
\begin{tabular}{l||c|c|c}
Number of events & 1 & 2 & 3 and more\\\hline
Time proportion & 86.86\%& 10.40\%& 2.74\%\\
Clip proportion &33.68\%& 17.71\%& 48.61\%\\
\end{tabular}
\caption{Proportion of overlapping events (test set)}
\label{tab:test_overlap}
\end{table}
The test set is designed such that the distribution in term of clips per class of sound event is similar to that of the weakly labeled training set. The size of the test set is such that it represents about 20\% of the size of the weakly labeled training set, it contains 288 clips (906 events). The test set is annotated with strong labels, with time boundaries (obtained by human annotators).

The minimum length for an event is 250~ms. The minimum duration of the pause between two events from the same class is 150~ms. When the silence between two consecutive events from the same class was less than 150~ms the events have been merged to a single event. The strong labels are provided in a tab-separated csv file under the following format:
\begin{center}
    {\sf [filename (str)][tab][onset (float)][tab][offset (float)][tab][class\_label (str)]}
\end{center}
The first column, is the name of the audio file downloaded from YouTube, the second column is the onset time in seconds, the third column is the offset time in seconds and the last column corresponds to the class of the sound event.

The amount of events per class is presented in Table~\ref{tab:dset}. This table also presents the mean and median duration of the events for each class. From these durations it is possible to categorize the events into three classes: short events (Alarm, cat, dishes, dog and speech), events with variable duration (blender and running water) and long events that span almost over the whole clip (electric shaver, frying and vacuum cleaner). This classification is confirmed by the observation of the duration distribution presented on Figure~\ref{dur_dist}.

One of the focus of this task is the development of approaches that can provide fine time-level segmentation while learning on weakly labeled data. The observation of the event duration distribution confirms that in order to perform well it is essential to design approaches that are efficient at detecting both short events and events that have a longer duration.

Table~\ref{tab:test_overlap} presents the time proportion of overlapping events in the test set. With overlapping events occurring about 12\% of the time it is important to design approaches that are able to detect and properly classify the overlapping events. However, this proportion is lower than the proportion of the clips containing multiple sound events (about 64\%). This tends to confirm that being able to discriminate between events that occur at different times within the clip is of a high importance as well.
This aspect reinforces the focus on efficient time segmentation.

\subsection{Evaluation set}
The evaluation set contains 880 10-seconds audio clips. The process to select the clips was similar to the process applied to select clips in the training set and the test set, in order to obtain a set with comparable classes distribution. Labels with time boundaries (obtained by human annotators) will be released after the DCASE 2018 challenge is concluded.

\section{Task description}
\label{sec:task}

The task consists of detecting sound events within web videos using weakly labeled training data. The detection within a 10-seconds clip should be performed with start and end timestamps. Task rules are detailed on the challenge webpage\footnote{\url{http://dcase.community/challenge2018/}}.
\subsection{Task evaluation}
Submissions will be evaluated with event-based measures for which the system output is compared to the reference labels event by event~\cite{mesaros_metrics_2016}. True positives are the occurrences when an event present in the system output corresponds to an event in the reference annotations. The correspondence between event boundaries are estimated with a 200ms collar tolerance on onsets and a tolerance on offsets that is the maximum of 200ms and 20\% of the event length. False positives are obtained when an event is present in the system output but not in the reference annotations (or not within the tolerance on the onset or the offset). False negatives are obtained when an event is present in the reference annotations but not in the system output (or not within the tolerance).

Submissions will be ranked according to the event-based F1-score. The F1-score is first computed class-wise over the whole evaluation set:
\begin{equation}
  F1_c = \frac{2TP_c}{2TP_c + FP_c + FN_c},
\end{equation}
where $TP_c$, $FP_c$ and $FN_c$ are the number of true positives, false positives and false negative for class of sound event $c$ over the whole evaluation set, respectively.

The final score is the F1-score average over class regardless of the number of events per class (macro-average):
\begin{equation}
  F1_{\mathrm{macro}} = \frac{\sum_{c\in \mathcal{C}} F1_c}{n_{\mathcal{C}}},
\end{equation}
where $\mathcal{C}$ is the classes ensemble and $n_{\mathcal{C}}$.

 Evaluation is done using sed\_eval toolbox~\cite{mesaros_metrics_2016}. The choice of the macro averaging allows for according the same importance to each class of sound event in an unbalanced scenario as in task 4.

\section{Baseline}
\label{sec:baseline}
\subsection{System description}
The baseline system is based on convolutional recurrent neural networks (CRNN). The number of 64 log mel-band magnitudes are extracted from 40~ms frames with 50\% overlap. Using these features, we train a first CRNN with three convolution layers (64 filters (3x3), max pooling (4) along the frequency axis and 30\% dropout), one recurrent layer (64 Gated Recurrent Units with 30\% dropout on the input), a dense layer (10 units sigmoid activation) and global average pooling across frames (Figure~\ref{baseline}).

\begin{figure}
  \centering
  \includegraphics[width=0.3\textwidth]{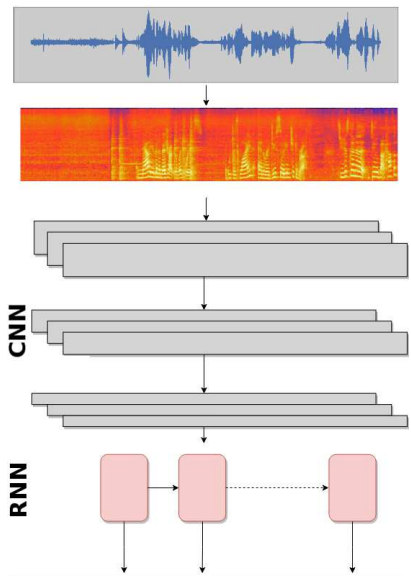}
  \caption{Baseline system description}
  \label{baseline}
\end{figure}

The system is trained for 100 epochs (early stopping after 15 epochs patience) on weak labels (`train/weak`, 1,578 clips). 20\% of this set is used for validation. This model is trained at clip level, the annotations only indicate if the event is present or not during the clip. The inputs are 500 frames long for a single output frame. This first model is used to predict labels of unlabeled files (`train/unlabel\_in\_domain`, 14,412 clips).

A second model based on the same architecture (Figure~\ref{baseline}) is trained on predictions of the first model on the unlabeled subset (`train/unlabel\_in\_domain`). The files with manual annotations (`train/weak`) are used for the validation of the model. The main difference with the first pass model is that the output dense layer is time-distributed in order to be able to predict events at the frame level. The inputs are 500 frames long, each of them labeled identically following clip labels. The model outputs a decision for each of these 500 frames. Median filtering over 51 frames ($\approx$ 1~s) is applied to the output of the network, in order to obtain the events onset and offset for each file. The full process is described in Algorithm~\ref{algo:script}.

\begin{algorithm}
  \centering
\begin{algorithmic}[1]
 \Procedure{Download the data (only the first time)}{}
 \EndProcedure
 \Procedure{First pass }{at clip level}
    \State Train a CRNN on weakly labeled data (`train/weak`)

    \Comment{20\% of data used for validation}
    \State Predict class for unlabeled data (`train/unlabel\_in\_domain`)
 \EndProcedure
 \Procedure{Second pass }{at frame level}
    \State Train a CRNN on labels predicted for the unlabeled data during the first pass (`train/unlabel\_in\_domain`)

      \Comment{weak data (`train/weak`) is used for validation~}

      \Comment{Labels are used at frames level. If a class is present in the clip, all frames contain the label.}
    \State Predict strong test labels (`test/`)

    \Comment{Predict an event with an onset and an offset}
    \EndProcedure
 \Procedure{Evaluation }{event based}
  \State Evaluate the prediction on the test set with respect to manual annotations
 \EndProcedure
 \caption{Script description \label{algo:script}}
 \end{algorithmic}
\end{algorithm}

\subsection{System performance}
The F1-score performance of the baseline system after the first pass (training on `train/weak` data) and after the second pass (training on `train/unlabel\_in\_domain` with labels obtained with the first pass system) are presented in Table~\ref{tab:fscore}. The second pass improves performance compared to the first pass in macro-average and for all classes expect for `Dishes`. Therefore, the baseline system takes advantage of the unlabeled data even though this could probably be done more efficiently by the systems submitted to the task.

The F1-score is close to zero for all the short event classes. It is slightly higher for event with variable duration (running water and blender) and performs the best for the long event class. This tends to indicate that the baseline performs very poorly in time-segmentation which is one of the focus of this task. Errors in segmentation are heavily penalized by the metric as they result in both a false positive and a false negative. Therefore, in order to improve performance, submitted systems should propose efficient time-segmentation approaches.
\begin{table}
  \centering
\begin{tabular}{l|c|c}
Class& 1$^{st}$ pass& 2$^{nd}$ pass\\
\hline\hline
Alarm/bell/ringing& 3.2\%&	3.9\%\\
Blender & 10.1\%&	15.4\%\\
Cat &0.0\%&0.0\%\\
Dishes &1.9\%&	0.0\%\\
Dog &0.0\%&0.0\%\\
Electric shaver/toothbrush &18.2\%&	32.4\%\\
Frying &9.4\%&	31.0\%\\
Running water&7.6\% &	11.4\%\\
Speech &0.0\%&0.0\%\\
Vacuum cleaner &24.8\%&	46.5\%\\

\hline\hline
Macro average& {\bf 7.51\%} &	{\bf 14.06\%}\\
\end{tabular}
\caption{Event based F1-scores}
\label{tab:fscore}
\end{table}

\section{conclusion}
\label{sec:conc}
This paper presents DCASE 2018 task 4 on large-scale weakly labeled semi-supervised SED in domestic environments. The goal is to exploit a small dataset of weakly labeled sound clips (without time boundaries) together with a larger unlabeled dataset to perform SED (with time boundaries). This indeed corresponds to a realistic scenario as obtaining time coded annotations is time consuming and is generally considered as one of the principal bottlenecks in training SED systems. The design of the dataset with a wide variability in event lengths and the choice of a metric that is heavily penalizing segmentation errors put a strong focus on the problem of localizing the events in time.

\bibliographystyle{IEEEtran}
\bibliography{refs}

\end{sloppy}
\end{document}